# On the relationship between the ellipticity of Galactic globular clusters and their X-ray luminosity

Georgi P. Petrov, Svetoslav Botev, Antoniya Valcheva & Petko Nedialkov

Department of Astronomy, Faculty of Physics, Sofia University "St. Kliment Ohridski",
5 J. Bourchier Blvd., Sofia 1164, Bulgaria

botev@phys.uni-sofia.bg

**Abstract.** We examine the dependence of the ellipticity of globular clusters in the Milky way on their X-ray luminosity using two modern catalogs and combine them with optical and X-ray data from the literature. Kolmogorov–Smirnov tests applied across multiple subsets reveal statistically significant differences in the ellipticity distributions when both $L_X$ and optical luminosity are considered. Two X-ray luminosity thresholds, $L_X^*(M_V = -7) = 10^{33.05}$ erg s$^{-1}$ and $L_X^*(M_V = -7) = 10^{32.01}$ erg s$^{-1}$, yield the most reliable distinction. In contrast to earlier findings based solely on optical data, our results demonstrate that globular clusters with the highest X-ray luminosity tend to have higher ellipticity on average.

**Key words:** Ellipticity, Globular clusters, Milky Way, X-ray luminosity

## Introduction

The apparent ellipticity is an important structural parameter of globular clusters (GCs). It describes their flattening and is defined as $\epsilon = (a-b)/a$ (Hubble, 1936), where $a$ and $b$ are the major and minor cluster axes. The apparent ellipticity is known to be related to their luminosity in the optical (van den Bergh, 2008) and near IR (Kostov et al., 2005). However, there is still a lack of convincing evidence that a similar relationship holds for the ellipticities of GCs and their X-ray luminosity.

Numerous observational and theoretical studies have established a strong connection between the internal dynamics of GCs and their populations of X-ray binaries. Clusters with higher central velocity dispersions exhibit enhanced dynamical interactions, leading to more frequent formation and hardening of binaries, as predicted by the Hills–Heggie law (Hills, 1975; Heggie, 1975). This is supported by recent Chandra surveys demonstrating that both the emissivity and abundance of X-ray sources correlate tightly with encounter rates, which scale with cluster velocity dispersion (Cheng et al., 2018a,b).

Studies of mass segregation in individual clusters, such as Terzan 5 and 47 Tucanae (Cheng et al., 2019a,b), further confirm that X-ray binaries preferentially concentrate toward the cluster core, consistent with models in which dynamical friction and interactions in dense, high-velocity-dispersion environments drive their production. Indeed, there is a variety of exotic objects observed in GCs like blue straggler stars (BSS), low-mass X-ray binaries (LMXBs), millisecond pulsars (MSPs), cataclysmic variables (CVs), and coronally active binaries (ABs). Some of them are very close binary systems (LMXBs, CVs, ABs), while others are remnants of close binaries (BSS, MSPs) formed and transformed through dynamical interactions of primordial binaries in GCs (see Cheng et al., 2019a; and the references therein). While direct studies of the relationship between ellipticity and X-ray binary population are limited, our study is focused on exploring a possible relationship between these two parameters.

Di Stefano (2002) was the first to probe whether six high X-ray luminosity clusters in M 31 galaxy have different ellipticities compared to 14 low X-ray luminosity clusters, but the limited sample size prevented definitive conclusions. Subsequent advances in X-ray studies of our neighboring galaxy allowed Kostov et al. (2005) to conclude, with better statistical support, that 23 clusters with high X-ray luminosity are more spherical than the 87 clusters with lower X-ray luminosity, at a 94% confidence level. The reliability of these results is limited by the fact that most of the clusters' flattening data used by Kostov et al. (2005) were taken from ground-based optical observations.

Botev et al. (2024) attempted a similar study for the Milky Way GCs but have not achieved conclusive results. They used two-samples Kolmogorov-Smirnov (K–S) test to show that high X-ray luminosity clusters have lower ellipticity than low X-ray luminosity clusters, but only at 80% confidence level. A significant issue was that eight of the 10 brightest X-ray Milky way GCs ($L_\mathrm{X} > 10^{34}$ erg s$^{-1}$) lack measured ellipticities, placing them among the 57 clusters with missing ellipticity values among 157 Milky way GCs in Harris (1996; 2010 edition). Recently, this obstacle was overcame by Cruz, Reyes & Anderson (2024) who homogeneously measured the ellipticities of 163 GCs using the third data release of the ESA mission Gaia (DR3).

This paper is structured as follows: Section 1 describes the data compilation, presenting the various catalogs used for ellipticity, X-ray luminosity, and other physical parameters of the GCs. This section also outlines the methodology for preparing the different samples used in our statistical analyses. In Section 2, we present and discuss the results, beginning with an analysis of the correlation between ellipticity and GCs characteristics, followed by the main results from the K–S tests, focusing on the relationship between ellipticity and the X-ray luminosity of the Galactic GCs. Finally, the Conclusion summarizes the key findings of this work.

# 1 Data

Table 1 presents our compilation list of 153 Milky Way GCs. The table contains ellipticities taken from Harris (1996; 2010 edition, hereafter H10) and Cruz, Reyes & Anderson (2024, hereafter CR24), as well as X-ray luminosities from Botev et al. (2024, hereafter B24) and Cheng et al. (2018a, hereafter C18). In the table, the second column presents absolute magnitudes ($M_\mathrm{V}$) for 152 GCs taken from H10 and the seventh column presents N-body simulation masses for 150 GCs taken from the 4th (March 2023) version of the GC database[1], compiled by Baumgardt et al. (hereafter BG23). The number sequence in the last column of Table 1 indicates some known evolutionary conditions of the listed GC, namely if it is collapsed, accreted, or has at least one peculiar feature like tidal tail, highly flattening, clumpy structure, or unreliable or noisy GC sequence.

---

[1] https://people.smp.uq.edu.au/HolgerBaumgardt/globular/

**Table 1.** Data for 153 Milky Way GCs, used for statistical analysis in this paper. Columns are as follows: (1) cluster ID, (2) $M_V$ absolute magnitude and (3) $\epsilon$ ellipticity from H10, (4) $\epsilon$ ellipticity from CR24, (5) $L_X([0.5-7]$ keV) X-ray luminosity within two core radii $2r_c$ adopted by B24, (6) $L_X([0.5-8]$ keV) within half light radius $r_h$ adopted by C18, (7) $M$ mass of the GC available in the 4th (Mar. 2023) version of the GC database compiled by BG23, (8) Features: core-collapsed [1] or normal GC [0] from H10, accreted [1] or in-situ [0] according to Belokurov & Kravtsov (2024), peculiar [1] (Grillmair et al., 1995; Lehmann & Scholz, 1997; Leon et al., 2000; Chen & Chen, 2010) or ordinary [0] (see the text).

| ID | $M_V$ (H10) mag | $\epsilon$ (H10) | $\epsilon$ (CR24) | $\log L_X$ (B24) $\mathrm{erg\,s^{-1}}$ | $\log L_X$ (C18) $\mathrm{erg\,s^{-1}}$ | $M$ (BG23) $M_\odot$ | Features indicator |
|---|---|---|---|---|---|---|---|
| (1) | (2) | (3) | (4) | (5) | (6) | (7) | (8) |
| NGC 104 | -9.42 | 0.09 | 0.059 | 33.534 | 32.884 | 8.53e+05 | 0/0/1 |
| NGC 288 | -6.75 | – | 0.021 | 32.766 | 32.107 | 9.60e+04 | 0/1/1 |
| NGC 362 | -8.43 | 0.01 | 0.035 | 32.914 | 32.964 | 2.52e+05 | 1/1/1 |
| Whiting 1 | -2.46 | – | 0.319 | – | – | 1.40e+03 | 0/1/0 |
| NGC 1261 | -7.80 | 0.07 | 0.042 | 32.719 | – | 1.72e+05 | 0/1/1 |
| Pal 1 | -2.52 | 0.22 | 0.2 | – | – | 9.30e+02 | 0/0/0 |
| AM 1 | -4.73 | – | 0.301 | – | – | 2.00e+04 | 0/1/0 |
| Eridanus | -5.13 | – | 0.31 | – | – | 9.30e+03 | 0/1/0 |
| Pal 2 | -7.97 | 0.05 | 0.127 | <30.5 | – | 2.20e+05 | 0/1/0 |
| NGC 1851 | -8.33 | 0.05 | 0.051 | 35.801 | – | 2.83e+05 | 0/1/1 |
| NGC 1904 | -7.86 | 0.01 | 0.027 | 32.957 | 32.870 | 1.80e+05 | 1/1/1 |
| NGC 2298 | -6.31 | 0.08 | 0.032 | 32.916 | – | 5.00e+04 | 0/1/1 |
| NGC 2419 | -9.42 | 0.03 | 0.073 | – | – | 7.80e+05 | 0/1/0 |
| Pyxis | -5.73 | – | 0.156 | – | – | 3.20e+04 | 0/1/0 |
| NGC 2808 | -9.39 | 0.12 | 0.031 | 36.092 | 33.461 | 7.91e+05 | 0/1/1 |
| E 3 | -4.12 | – | 0.112 | 32.131 | – | 2.60e+03 | 0/0/0 |
| Pal 3 | -5.69 | – | 0.221 | – | – | 1.90e+04 | 0/1/0 |
| NGC 3201 | -7.45 | 0.12 | 0.065 | 32.168 | <32.173 | 1.93e+05 | 0/1/1 |
| Pal 4 | -6.01 | – | 0.226 | – | – | 1.50e+04 | 0/1/0 |
| NGC 4147 | -6.17 | 0.08 | 0.08 | – | – | 4.50e+04 | 0/1/0 |
| NGC 4372 | -7.79 | 0.15 | 0.021 | 32.810 | – | 1.90e+05 | 0/0/1 |
| Rup 106 | -6.35 | – | 0.057 | – | – | 3.40e+04 | 0/1/0 |
| NGC 4590 | -7.37 | 0.05 | 0.062 | 32.775 | – | 1.30e+05 | 0/1/1 |
| NGC 4833 | -8.17 | 0.07 | 0.022 | – | – | 1.86e+05 | 0/0/0 |
| NGC 5024 | -8.71 | 0.01 | 0.039 | 33.032 | 33.083 | 5.02e+05 | 0/1/1 |
| NGC 5053 | -6.76 | 0.21 | 0.137 | – | – | 6.30e+04 | 0/1/1 |
| NGC 5139 | -10.26 | 0.17 | 0.064 | 33.246 | 33.149 | 3.94e+06 | 0/0/1 |
| NGC 5272 | -8.88 | 0.04 | 0.025 | 33.774 | 32.961 | 4.09e+05 | 0/1/1 |
| NGC 5286 | -8.74 | 0.12 | 0.047 | 32.734 | 33.130 | 4.24e+05 | 1/1/1 |
| NGC 5466 | -6.98 | 0.11 | 0.102 | <30.5 | – | 5.60e+04 | 0/1/1 |
| NGC 5634 | -7.69 | 0.02 | 0.068 | – | – | 2.50e+05 | 0/1/0 |
| NGC 5694 | -7.83 | 0.04 | 0.093 | – | – | 2.70e+05 | 0/1/1 |
| IC 4499 | -7.32 | 0.08 | 0.061 | – | – | 1.50e+05 | 0/1/0 |
| NGC 5824 | -8.85 | 0.03 | 0.061 | <30.5 | <33.444 | 7.50e+05 | 0/1/1 |
| Pal 5 | -5.17 | – | 0.13 | 33.949 | – | 1.30e+04 | 0/1/1 |
| NGC 5897 | -7.23 | 0.08 | 0.045 | – | – | 1.70e+05 | 0/0/1 |
| NGC 5904 | -8.81 | 0.14 | 0.057 | 32.530 | 32.704 | 3.92e+05 | 0/1/1 |
| NGC 5927 | -7.81 | 0.04 | 0.019 | 32.020 | 32.687 | 2.93e+05 | 0/0/1 |
| NGC 5946 | -7.18 | 0.16 | 0.057 | 31.656 | <32.675 | 1.10e+05 | 1/0/0 |
| BH 176 | -4.06 | – | 0.133 | – | – | – | 0/–/0 |
| NGC 5986 | -8.44 | 0.06 | 0.04 | – | – | 2.99e+05 | 0/1/0 |
| Lynga 7 | -6.60 | – | 0.046 | – | – | 6.80e+04 | 0/0/0 |
| Pal 14 | -4.80 | – | 0.143 | – | – | 1.90e+04 | 0/1/0 |

**Table 1.** (continuing from the previous page)

| (1) | (2) | (3) | (4) | (5) | (6) | (7) | (8) |
|---|---|---|---|---|---|---|---|
| NGC 6093 | -8.23 | 0 | 0.035 | 33.234 | 33.375 | 3.21e+05 | 0/0/0 |
| NGC 6121 | -7.19 | 0 | 0.018 | 32.172 | 32.721 | 9.04e+04 | 0/0/1 |
| NGC 6101 | -6.94 | 0.05 | 0.066 | 33.021 | – | 1.70e+05 | 0/1/1 |
| NGC 6144 | -6.85 | 0.25 | 0.037 | 32.950 | 32.695 | 8.50e+04 | 0/0/0 |
| NGC 6139 | -8.36 | 0.05 | 0.041 | 33.035 | 33.396 | 3.50e+05 | 0/0/0 |
| Terzan 3 | -4.82 | – | 0.127 | 32.318 | <32.029 | 3.30e+04 | 0/0/0 |
| NGC 6171 | -7.12 | 0.02 | 0.019 | 32.340 | – | 6.12e+04 | 0/0/0 |
| NGC 6205 | -8.55 | 0.11 | 0.05 | 32.920 | 32.961 | 4.84e+05 | 0/0/1 |
| NGC 6229 | -8.06 | 0.05 | 0.14 | – | – | 2.00e+05 | 0/1/1 |
| NGC 6218 | -7.31 | 0.04 | 0.017 | 32.408 | 32.603 | 1.06e+05 | 0/0/1 |
| FSR 1735 | -6.45 | – | 0.43 | – | – | 1.00e+05 | 0/0/0 |
| NGC 6235 | -6.29 | 0.13 | 0.036 | – | – | 9.60e+04 | 0/0/0 |
| NGC 6254 | -7.48 | 0 | 0.025 | – | – | 1.89e+05 | 0/0/1 |
| NGC 6256 | -7.15 | – | 0.091 | 33.019 | 33.468 | 1.11e+05 | 1/0/0 |
| Pal 15 | -5.51 | – | 0.113 | – | – | 5.30e+04 | 0/1/0 |
| NGC 6266 | -9.18 | 0.01 | 0.052 | 33.652 | 33.806 | 5.81e+05 | 1/0/1 |
| NGC 6273 | -9.13 | 0.27 | 0.049 | 32.705 | – | 7.20e+05 | 0/0/1 |
| NGC 6284 | -7.96 | 0.03 | 0.058 | – | – | 1.70e+05 | 1/0/0 |
| NGC 6287 | -7.36 | 0.13 | 0.073 | 32.715 | 32.989 | 1.30e+05 | 0/0/0 |
| NGC 6293 | -7.78 | 0.03 | 0.055 | 32.243 | <33.004 | 1.40e+05 | 1/0/0 |
| NGC 6304 | -7.30 | 0.02 | 0.053 | 32.585 | 33.236 | 1.00e+05 | 0/0/0 |
| NGC 6316 | -8.34 | 0.04 | 0.095 | – | – | 3.50e+05 | 0/0/0 |
| NGC 6341 | -8.21 | 0.1 | 0.042 | 32.631 | 32.734 | 2.73e+05 | 0/1/1 |
| NGC 6325 | -6.96 | 0.12 | 0.046 | 31.815 | 32.501 | 6.20e+04 | 1/0/0 |
| NGC 6333 | -7.95 | 0.04 | 0.046 | 32.011 | <32.711 | 3.10e+05 | 0/0/1 |
| NGC 6342 | -6.42 | 0.18 | 0.053 | 31.149 | 32.928 | 3.77e+04 | 1/0/1 |
| NGC 6356 | -8.51 | 0.03 | 0.073 | – | – | 5.70e+05 | 0/0/0 |
| NGC 6355 | -8.07 | – | 0.052 | 31.824 | 32.862 | 9.91e+04 | 1/0/0 |
| NGC 6352 | -6.47 | 0.07 | 0.031 | – | 32.617 | 6.00e+04 | 0/0/0 |
| IC 1257 | -6.15 | – | 0.144 | – | – | 1.80e+04 | 0/1/0 |
| Terzan 2 | -5.88 | – | 0.174 | 35.201 | – | 8.00e+04 | 1/0/0 |
| NGC 6366 | -5.74 | 0.16 | 0.03 | 32.210 | 32.501 | 4.70e+04 | 0/0/1 |
| Terzan 4 | -4.48 | – | 0.284 | <30.5 | – | 1.80e+05 | 0/0/0 |
| HP 1 | -6.46 | – | 0.082 | – | – | 1.40e+05 | 1/0/0 |
| NGC 6362 | -6.95 | 0.07 | 0.028 | 32.514 | <32.589 | 1.17e+05 | 0/0/1 |
| Liller 1 | -7.32 | – | 0.253 | 35.311 | – | 1.00e+06 | 0/0/0 |
| NGC 6380 | -7.50 | – | 0.072 | – | – | 3.41e+05 | 1/0/1 |
| Terzan 1 | -4.41 | – | 0.089 | <30.5 | 33.389 | 2.00e+05 | 1/0/0 |
| Ton 2 | -6.17 | – | 0.141 | – | – | 4.30e+04 | 0/0/0 |
| NGC 6388 | -9.41 | 0.01 | 0.035 | 33.858 | 33.948 | 1.31e+06 | 0/0/1 |
| NGC 6402 | -9.10 | 0.11 | 0.023 | – | 33.248 | 6.00e+05 | 0/0/1 |
| NGC 6401 | -7.90 | 0.15 | 0.1 | 32.748 | <33.366 | 1.21e+05 | 0/0/0 |
| NGC 6397 | -6.64 | 0.07 | 0.03 | 32.598 | 33.124 | 8.24e+04 | 1/0/1 |
| Pal 6 | -6.79 | – | 0.119 | – | – | 8.60e+04 | 0/0/0 |
| NGC 6426 | -6.67 | 0.15 | 0.094 | – | – | 7.00e+04 | 0/1/0 |
| Djorg 1 | -6.98 | – | 0.155 | – | – | 8.40e+04 | 0/0/0 |
| Terzan 5 | -7.42 | – | 0.114 | 36.251 | 34.068 | 1.10e+06 | 0/0/0 |
| NGC 6440 | -8.75 | 0.01 | 0.051 | 34.110 | 33.860 | 5.70e+05 | 0/0/0 |
| NGC 6441 | -9.63 | 0.02 | 0.049 | 36.206 | – | 1.39e+06 | 0/0/0 |
| Terzan 6 | -7.59 | – | 0.259 | 34.847 | – | 1.00e+05 | 1/0/0 |
| NGC 6453 | -7.22 | 0.09 | 0.1 | 33.393 | 32.710 | 1.68e+05 | 1/0/0 |
| UKS 1 | -6.91 | – | 0.324 | – | – | 8.00e+04 | 0/0/0 |
| NGC 6496 | -7.20 | 0.16 | 0.041 | – | – | 7.40e+04 | 0/0/1 |
| Terzan 9 | -3.71 | – | 0.226 | 32.761 | 33.378 | 1.40e+05 | 1/0/0 |
| Djorg 2 | -7.00 | – | 0.183 | 33.271 | – | 1.30e+05 | 0/0/0 |

71

**Table 1.** (continuing from the previous page)

| (1) | (2) | (3) | (4) | (5) | (6) | (7) | (8) |
|---|---|---|---|---|---|---|---|
| NGC 6517 | -8.25 | 0.06 | 0.058 | 32.075 | <32.656 | 2.20e+05 | 0/0/0 |
| Terzan 10 | -6.35 | – | 0.238 | – | – | 3.00e+05 | 0/0/0 |
| NGC 6522 | -7.65 | 0.06 | 0.07 | 32.473 | <33.004 | 2.10e+05 | 1/0/0 |
| NGC 6535 | -4.75 | 0.08 | 0.041 | 31.611 | <32.061 | 2.00e+04 | 0/0/0 |
| NGC 6528 | -6.57 | 0.11 | 0.092 | 32.328 | 32.572 | 9.40e+04 | 0/0/0 |
| NGC 6539 | -8.29 | 0.08 | 0.037 | 33.014 | 33.501 | 2.20e+05 | 0/0/0 |
| NGC 6540 | -6.35 | – | 0.12 | 32.309 | – | 5.60e+04 | 0/0/0 |
| NGC 6544 | -6.94 | 0.22 | 0.142 | 30.613 | 32.233 | 8.10e+04 | 1/0/0 |
| NGC 6541 | -8.52 | 0.12 | 0.095 | 33.240 | 33.354 | 2.57e+05 | 1/0/1 |
| 2MS-GC01 | -6.11 | – | – | 30.980 | – | 4.10e+04 | 0/0/0 |
| ESO 280-SC06 | -4.87 | – | 0.135 | – | – | 4.00e+04 | 0/1/0 |
| NGC 6553 | -7.77 | 0.17 | 0.032 | 32.547 | 32.998 | 2.30e+05 | 0/0/0 |
| 2MS-GC02 | -4.86 | – | – | 32.303 | – | 1.60e+04 | 0/0/0 |
| NGC 6558 | -6.44 | – | 0.14 | <30.5 | <33.101 | 3.10e+04 | 1/0/0 |
| IC 1276 | -6.67 | – | 0.065 | – | – | 7.40e+04 | 0/0/1 |
| Terzan 12 | -4.14 | – | 0.221 | – | – | 3.80e+04 | 0/0/0 |
| NGC 6569 | -8.28 | 0 | 0.061 | 32.698 | <32.880 | 2.30e+05 | 0/0/1 |
| BH 261 | -4.19 | 0.03 | 0.146 | – | – | 2.40e+04 | 0/0/0 |
| GLIMPSE02 | - | – | – | 32.772 | – | – | 0/–/0 |
| NGC 6584 | -7.69 | – | 0.041 | – | – | 1.10e+05 | 0/1/0 |
| NGC 6624 | -7.49 | 0.06 | 0.058 | <30.5 | – | 1.03e+05 | 1/0/0 |
| NGC 6626 | -8.16 | 0.16 | 0.065 | 35.795 | 33.766 | 2.70e+05 | 0/0/0 |
| NGC 6638 | -7.12 | 0.01 | 0.079 | 32.962 | 33.167 | 1.20e+05 | 0/0/1 |
| NGC 6637 | -7.64 | 0.01 | 0.035 | 33.098 | 33.220 | 1.38e+05 | 0/0/1 |
| NGC 6642 | -6.66 | 0.03 | 0.1 | 31.621 | <32.714 | 3.90e+04 | 1/0/0 |
| NGC 6652 | -6.66 | 0.2 | 0.039 | 33.336 | – | 4.10e+04 | 0/0/0 |
| NGC 6656 | -8.50 | 0.14 | 0.056 | 32.503 | 32.672 | 4.70e+05 | 0/0/1 |
| Pal 8 | -5.51 | – | 0.106 | – | – | 7.10e+04 | 0/0/0 |
| NGC 6681 | -7.12 | 0.01 | 0.027 | 31.652 | 32.140 | 1.05e+05 | 1/0/1 |
| GLIMPSE01 | -5.91 | – | – | 32.832 | 33.193 | – | 0/–/0 |
| NGC 6712 | -7.50 | 0.11 | 0.054 | 35.924 | – | 9.50e+04 | 0/0/0 |
| NGC 6715 | -9.98 | 0.06 | 0.02 | 33.604 | 34.093 | 1.59e+06 | 0/1/1 |
| NGC 6717 | -5.66 | 0.01 | 0.038 | 33.691 | 32.985 | 2.60e+04 | 0/0/0 |
| NGC 6723 | -7.83 | 0 | 0.03 | – | – | 1.97e+05 | 1/0/0 |
| NGC 6749 | -6.70 | – | 0.064 | – | – | 2.00e+05 | 0/0/0 |
| NGC 6752 | -7.73 | 0.04 | 0.023 | 32.581 | 33.045 | 2.61e+05 | 1/0/1 |
| NGC 6760 | -7.84 | 0.04 | 0.063 | 32.122 | <32.481 | 2.90e+05 | 0/0/0 |
| NGC 6779 | -7.41 | 0.03 | 0.036 | – | – | 1.70e+05 | 0/1/0 |
| Terzan 7 | -5.01 | – | 0.107 | – | – | 2.20e+04 | 0/1/0 |
| Pal 10 | -5.79 | – | 0.115 | – | <32.549 | 1.30e+05 | 0/0/0 |
| Arp 2 | -5.29 | – | 0.12 | – | – | 3.90e+04 | 0/1/0 |
| NGC 6809 | -7.57 | 0.02 | 0.015 | 32.588 | 32.100 | 1.97e+05 | 0/0/1 |
| Terzan 8 | -5.07 | – | 0.149 | – | – | 7.60e+04 | 0/1/0 |
| Pal 11 | -6.92 | – | 0.069 | – | – | 1.00e+04 | 0/0/0 |
| NGC 6838 | -5.61 | 0 | 0.016 | 32.328 | 32.427 | 3.80e+04 | 0/0/1 |
| NGC 6864 | -8.57 | 0.07 | 0.052 | – | – | 4.60e+05 | 0/1/1 |
| NGC 6934 | -7.45 | 0.01 | 0.031 | – | – | 1.50e+05 | 0/1/1 |
| NGC 6981 | -7.04 | 0.02 | 0.034 | – | – | 8.10e+04 | 0/1/1 |
| NGC 7006 | -7.67 | 0.01 | 0.117 | – | – | 1.30e+05 | 0/1/0 |
| NGC 7078 | -9.19 | 0.05 | 0.039 | 33.780 | – | 5.18e+05 | 1/0/1 |
| NGC 7089 | -9.03 | 0.11 | 0.027 | 32.722 | 33.152 | 6.24e+05 | 0/1/1 |
| NGC 7099 | -7.45 | 0.01 | 0.032 | 31.820 | 32.940 | 1.21e+05 | 1/0/0 |
| Pal 12 | -4.47 | – | 0.143 | – | – | 6.20e+03 | 0/1/1 |
| Pal 13 | -3.76 | – | 0.21 | – | – | 2.80e+03 | 0/1/0 |
| NGC 7492 | -5.81 | 0.24 | 0.084 | <30.5 | – | 2.00e+04 | 0/1/1 |

## 1.1 Milky way GC X-ray luminosities and ellipticities

The main source of measurements of the X-ray luminosity of the GCs that is referred to in this paper is the compilation of B24 which provides 89 $L_X([0.5–7]$ keV), integrated within $2r_c$ around the center of each cluster. The individual X-ray fluxes $F_X$ of all detected resolved sources were taken from The Chandra Source Catalog (CSC), Release 2.0 (Evans et al., 2010–2019), XMM-Newton Serendipitous Source Catalog 4XMM-DR9 (Webb et al., 2020) and 2SXPS Swift X-ray telescope point source catalog (Evans et al., 2020). The total number of accounted individual sources was more than 120 in the cases of the NGC 104 and NGC 5139. All fluxes have been converted to the Chandra range of photon energies 0.5 to 7 keV and averaged over missions. The final X-ray luminosities were calculated by taking into account not only the distance but also the absorption of X-ray photons by neutral hydrogen atoms along the line of sight, assuming an efficient energy of 2.3 keV. The column density of hydrogen $N_H$ was obtained from the color excess E(B–V) of a cluster, given in the catalog of H10, assuming a standard gas-to-dust ratio (Bohlin et al., 1978). There are eight low X-ray luminosity clusters among 89 GCs in B24 with no individual sources detected within $2r_c$ for which an upper limit of $L_X([0.5–7]$ keV)=30.5 keV was tentatively ascribed (see Table 1).

The second source (C18) for $L_X([0.5–8]$ keV) luminosity of the 69 Milky way GCs is based on purely archival Chandra images and focused on the weak individual X-ray sources, mainly cataclysmic variables (CVs) and coronally active binaries (ABs) located within half-light circle, or radius $r_h$ from the cluster center. The unabsorbed fluxes were calculated assuming a power-law model with a photon index of 2.0. Although C18 claimed that all GCs hosting luminous LMXB sources were removed from their sample, we still find three GCs, namely Terzan 1, NGC 6640 and Terzan 5, from the LMXB catalog of Liu et al. (2007) among their 69 entries.

The difference between the half light radius $L_X([0.5–8]$ keV) luminosity adopted by C18 and the two core radii $L_X([0.5–7]$ keV) luminosity adopted by B24 as a function of $L_X([0.5–7]$ keV) luminosity adopted by B24 is given in logarithmic scale in Fig. 1. In general, for 58 of 66 GCs in common the agreement is around or less than one order of magnitude, but most of the GCs in C18 are found above the $\Delta \log L_X=0$ line on the plot since the number of the individual X-ray sources within $r_h$ is greater than the sources within $2r_c$ which encompass smaller area and $L_X([0.5–8]$ keV) luminosity also covers a wider range of photon energies. Nevertheless, one should keep in mind that 15 among 58 X-ray GCs luminosities (C18) are actually upper limits which acts towards smaller differences $\Delta \log L_X$ in Fig. 1.

Differences of $\Delta \log L_X \sim 3$ (see Fig. 1) occur when the brightest X-ray source is located outside the $2r_c$ radius but still within $r_h$ of the GC. For instance, GC Terzan 1 with $r_h = 230''$ and $2r_c = 5''$ contains a bright source at $r = 42''$, which in B24 was not accounted for, explaining its lower $L_X$ value in their list. At the high-energy end of the X-ray luminosity in Fig. 1, we find GCs like NGC 6640 with $r_h = 29''$ and $2r_c = 17''$, containing a bright LMXB source at $r = 9''$. This source was apparently excluded in C18, resulting in a lower $L_X([0.5–8]$ keV) luminosity. A similar situation applies to Terzan 5 with $r_h = 43''$ and $2r_c = 19''$, where a very bright LMXB source observed at $r = 6''$ was also eliminated in C18. Since most of the brightest sources are highly variable and exhibit quiescent phases from time to time, we cannot rule out the possibility that they were indeed in such a state during the Chandra observations analyzed by C18.

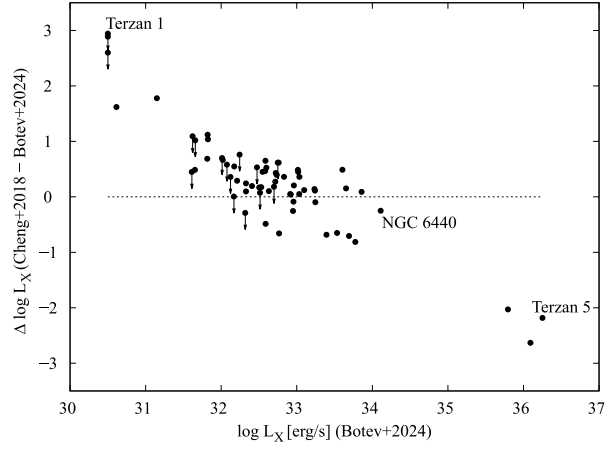

**Fig. 1.** The difference between the half light radius $L_X$([0.5–8] keV) luminosity adopted by C18 and the two core radii $L_X$([0.5–7] keV) luminosity adopted by B24 as a function of $L_X$([0.5–7] keV). Note that 58 among 66 all GCs have luminosities matching within an order of magnitude and 15 GCs luminosities (C18) are actually upper limits, indicated with arrows.

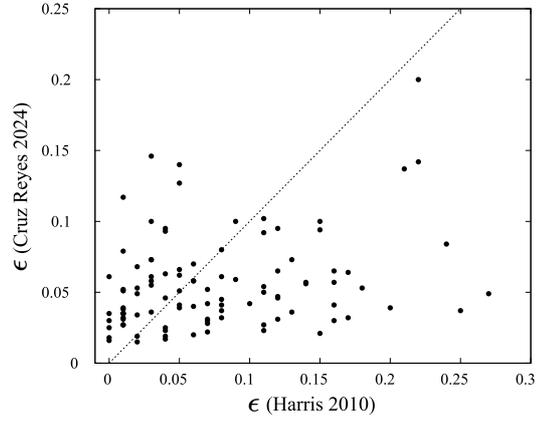

**Fig. 2.** Comparison of 100 GCs ellipticity $\epsilon$ estimates by CR24 vs. those derived from H10.

We use the catalogs of H10 and CR24 for the structural parameter ellipticity $\epsilon$ of the Milky Way GCs (see Table 1). The first source contains data for 100 and the second - for 149 GCs. For 100 GCs in common, the K–S test confirms their statistical difference in ellipticity at significance levels of 99.0% and 93.2% – for the subsamples with available X-ray luminosity. Figure 2 presents comparison of ellipticities $\epsilon$ estimated by CR24 and those of H10. The weak positive correlation coefficients are only $0.28 \pm 0.09$ between the sample of 100 GCs and $0.30 \pm 0.09$ for the sub-sample of 73 GCs with X-ray data. Outlined above weak correlations based on different ellipticity sources, namely H10 and CR24, prevented us from making a convincing conclusion about ellipticity differences between high and low X-ray luminosity clusters.

## 1.2  K–S tests samples preparation

We used data from Table 1 to calculate the ratio mass-to-$L_X$ of the GCs. For these purposes, we converted the X-ray luminosity into total solar luminosity assuming $\log L_\odot = 33.5827$ (Allen, 1973). Thus, we were able to analyze six different relationships between the two systems of ellipticities and three global characteristics of GCs like absolute magnitude $M_V$, X-ray luminosity $L_X$ and mass-to-$L_X$ ratio.

By combining available data for ellipticity and X-ray luminosity $L_X$ and applying additional selection criteria, we defined 12 major samples used to perform K–S two-sample tests for GCs ellipticities higher and lower than a certain threshold $L_X^*$. When that threshold was also varied so that the minimum number of objects $n_{min}$ in each compared sample is between 7 and 15 the performed K–S tests are 538 in total. The summarized information about the K–S test samples is given in Table 2.

**Table 2.** The properties of the data samples used to perform K–S tests for statistically different ellipticity distributions conditioned by a fixed threshold $L_X^*$. The abbreviations are the same as in Table 1. Columns are as follows: (1) Sample number, (2) $\epsilon$ GCs' ellipticity source, (3) $L_X$ X-ray luminosity source, (4) $n$ the number of GCs with both ellipticity and X-ray data, (5) Additional selection criterion, (6) $n_{KS}$ the total number of K–S tests performed with corresponding sample, (7) $n_{min}$ the minimum number of points at the extreme sides of the threshold X-ray luminosity $L_X^*$, (8) Shown on figure No.

| Sample | $\epsilon$ | $L_X$ | $n$ | Additional selection criterion | $n_{KS}$ | $n_{min}$ | Figure |
|--------|------------|-------|-----|-------------------------------|----------|-----------|--------|
| (1) | (2) | (3) | (4) | (5) | (6) | (7) | (8) |
| 1 | H10 | B24 | 73 | none | 54 | 10 | 5a |
| 2 | CR24 | B24 | 89 | none | 60 | 15 | 5a |
| 3 | H10 | B24 | 73 | $L_X^*(M_V = -7)$ | 58 | 8 | 5b |
| 4 | CR24 | B24 | 89 | $L_X^*(M_V = -7)$ | 70 | 10 | 5b |
| 5 | H10 | C18 | 59 | none | 40 | 10 | 5c |
| 6 | CR24 | C18 | 68 | none | 49 | 10 | 5c |
| 7 | H10 | B24 | 55 | non-collapsed | 36 | 10 | 6a |
| 8 | CR24 | B24 | 64 | non-collapsed | 45 | 10 | 6a |
| 9 | H10 | B24 | 53 | 'in situ' | 34 | 10 | 6b |
| 10 | CR24 | B24 | 67 | 'in situ' | 48 | 10 | 6b |
| 11 | H10 | B24 | 29 | ordinary | 16 | 7 | 6c |
| 12 | CR24 | B24 | 43 | ordinary | 28 | 8 | 6c |

Figure 3 illustrates an example of our approach to construct two ellipticity subsamples formed by choosing different thresholds $L_X^*$ when both log $L_X$ and $M_V$ are taken into consideration (see Table 2, sample 4). As seen in that figure, most of the objects are situated along a solid line in an interval between the upper and the lower dotted-dashed borders, with the exception of two groups of 11 objects each, respectively, above and below the considered region. The solid line through the central group of points is a linear regression with a slope of $-0.3441$. The intercept of that line was moved so that different GCs are left above and below the line. Then, the value of $L_X$ erg s$^{-1}$ at $M_V = -7$ mag for such lines is considered as a threshold indicator $L_X^*$ that defines the dividing line for GCs and their corresponding ellipticities.

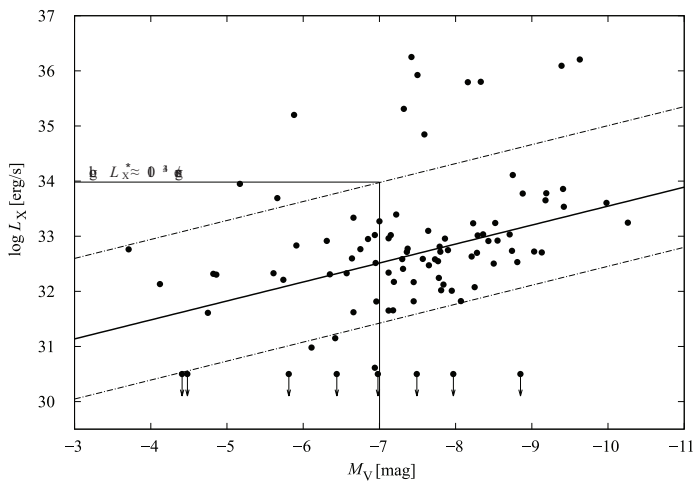

**Fig. 3.** Log $L_X$ (B24) vs $M_V$ plot for 89 GCs (see Table 1). Three examples of dividing lines, used to select pairs of samples for K–S tests are shown. The bulk of the GCs are situated along the solid line in an interval between the upper and the lower dotted-dashed lines. These lines illustrate two different pairs, corresponding to different thresholds. The upper one, with the indicated threshold $L_X^* \approx 10^{34}$ erg s$^{-1}$ at $M_V = -7$ mag divides the 89 GCs into a sample with 11 objects with $L_X$ higher than $L_X^*$ and sample with 78 objects with $L_X$ lower than $L_X^*$. In contrast, the lower dotted-dashed line forms another pair of 78 objects with $L_X$, higher and 11 objects with $L_X$ lower than the other $L_X^*$.

## 2 Results and Discussion

### 2.1 Some global clusters characteristics and observed GCs flattening

As we stated above the ellipticities published by CR24 and H10 do not show statistically significant correlation (Fig. 2). This motivated us to look for any relationship between the ellipticities of clusters and some global parameters using the two catalogs in a separate way. This is demonstrated in Fig. 4. The ellipticities in the left diagrams in Fig. 4 have been taken from H10 and in the right diagrams – from CR24. In the two (a) panels, the ellipticity $\epsilon$ is plotted versus the absolute magnitude $M_V$. In a similar diagram Davoust & Prugniel (1990) found a correlation between ellipticity and absolute magnitude $M_V$ for GCs in our Galaxy and in M31. According to them, the luminous GCs are, on average, rounder. Later, van den Bergh (2008) confirmed their result for 92 Galactic GCs and the fact that the faintest ($M_V > -7$ mag) are the flattest ones. He conducted a K–S test, which indicated a 97% probability of a statistically significant difference between the two subsamples. Chen & Chen (2010) also provided evidence supporting the hypothesis that brighter (and probably more massive) Milky Way GCs tend to be rounder. Our results confirm this tendency, but it is more clear in the right (a) panel, where the ellipticities from CR24 have been adopted. Note that this diagram includes 49 additional objects and seven more ellipticity values for objects with $M_V > -7$ mag, compared to the left (a) panel. Our K–S tests revealed differences of 99.8% and 100% between the $\epsilon$ distributions of the brightest objects ($M_V < -7$ mag) and the faintest ones, based on $\epsilon$ estimates from H10 and CR24, respectively. Moreover, when only $\epsilon$ of CR24 is considered, K–S tests show that regardless of the threshold $M_V$ there is always a statistically significant difference between the two subsamples. In addition, we calculated the Pearson's correlation coefficient $R$ between $\epsilon$ and $M_V$. For all 100 GCs from H10 it is $R = 0.19 \pm 0.10$ and for the 73 GCs with available X-ray luminosity $R = 0.09 \pm 0.10$. Better correlations have been achieved in the GC subsamples based on the ellipticities reported by CR24: for all 149 GCs, $R = 0.55 \pm 0.07$, and for the 89 X-ray objects, $R = 0.42 \pm 0.08$.

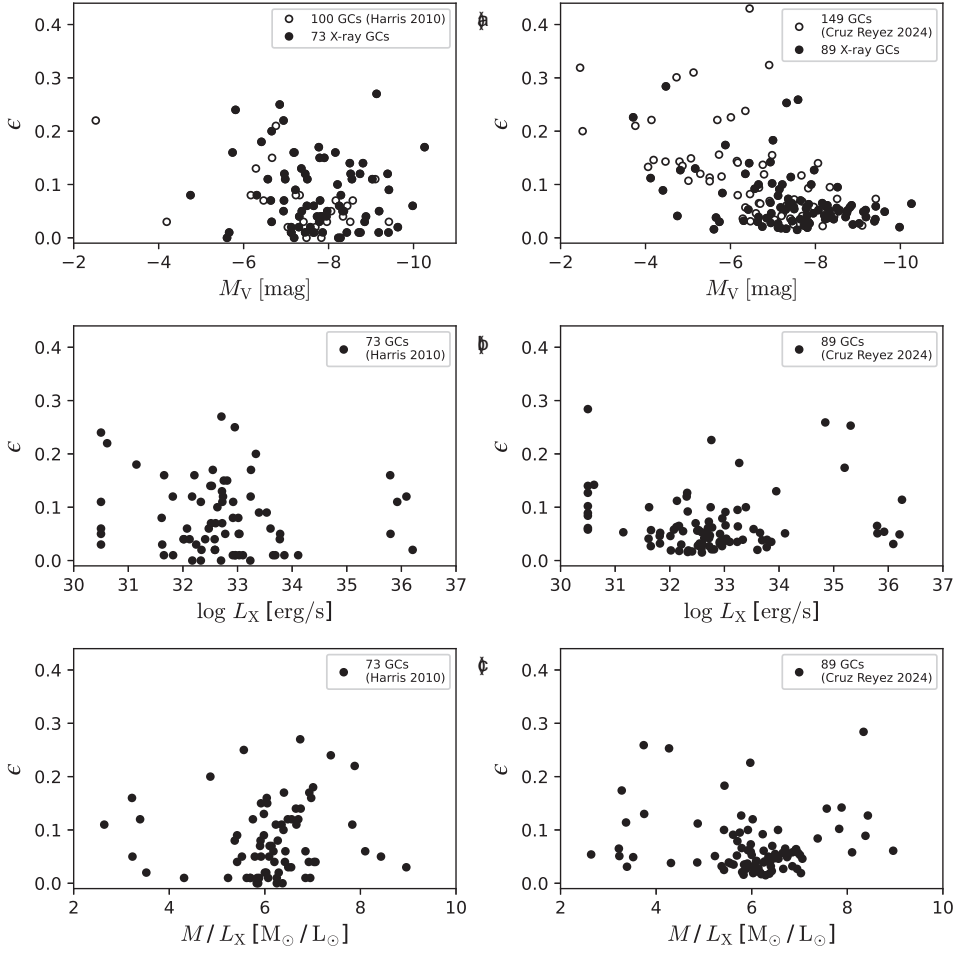

**Fig. 4.** Ellipticity vs. fundamental GC parameters: absolute magnitude $M_V$, X-ray luminosity $L_X$, and the GC mass-to-$L_X$ ratio. The ellipticity data in the left panels is sourced from H10, while the right panels utilize measurements from CR24. GCs with available $L_X$ data, as reported by B24, are indicated by filled black circles, while the open circles represent GCs without $L_X$. Additional details can be found in Table 1.

The two (b) panels in Fig. 4 represent $\epsilon$ versus $\log L_X$. In both the two distinct groups of objects are visible (see Sarazin et al., 1999): GCs containing dim X-ray sources ($\log L_X < 34.5$ [erg s$^{-1}$]) and high-luminosity X-ray GCs ($\log L_X > 34.5$ [erg s$^{-1}$]), but in the right (b) panel, based on CR24, 74 GCs have $\epsilon \leq 0.1$. However, no correlation has been found between $\epsilon$ and $L_X$ in these diagrams.

In the two (c) panels in the same figure, $\epsilon$ is plotted versus the GC mass-to-$L_X$ ratio. About 80% of the GCs have $5 \lesssim M / L_X \lesssim 7$ values in both diagrams and on the right side of this region, are located most of the GCs that are placed below the lower dashed line in Fig. 3 (with lower $L_X$ values) and on the left side are located mainly higher $L_X$ objects (above the upper dashed line in Fig. 3). Of course, there is no complete correspondence between these objects in the two figures, as the mass of the GCs is estimated mainly by N-body simulations rather than by assuming a fixed mass-to-luminosity ratio in the optical. Also, there is no evidence for $\epsilon - M / L_X$ correlation in both diagrams.

## 2.2   Cluster's ellipticity – X-ray luminosity relationship, K–S tests

We varied the X-ray luminosity threshold $L_X^*$ proposed by B24 (denoted in that paper as $L_X^{lim}$ on p. 73) in order to produce appropriate pairs of data sets accounting for their ellipticities and to compare their cumulative distributions via K–S test. A given pair of data sets contains a set of $n_1$ ellipticities for objects with $L_X < L_X^*$ and $n_2$ ellipticities for objects with $L_X \geq L_X^*$. We calculated the probability $(1-p)$ or the level of significance $P_{KS}$ for statistically different ellipticity $\epsilon$ distributions of the sets and plotted it as function of the X-ray luminosity threshold $L_X^*$ (see Fig. 4 in the above cited paper where the Y-axis presents the K–S test probability $p$ for identical distributions).

The 538 test results, whose total number equals the sum of $n_{KS}$ (see Table 2, col. 6) are presented graphically in Fig. 5 (see Table 2, data samples 1–6) and Fig. 6 (see Table 2, data samples 7–12) where the broken solid lines connect the K–S results for the CR24 ellipticities and the broken dashed lines – for the ellipticities from H10.

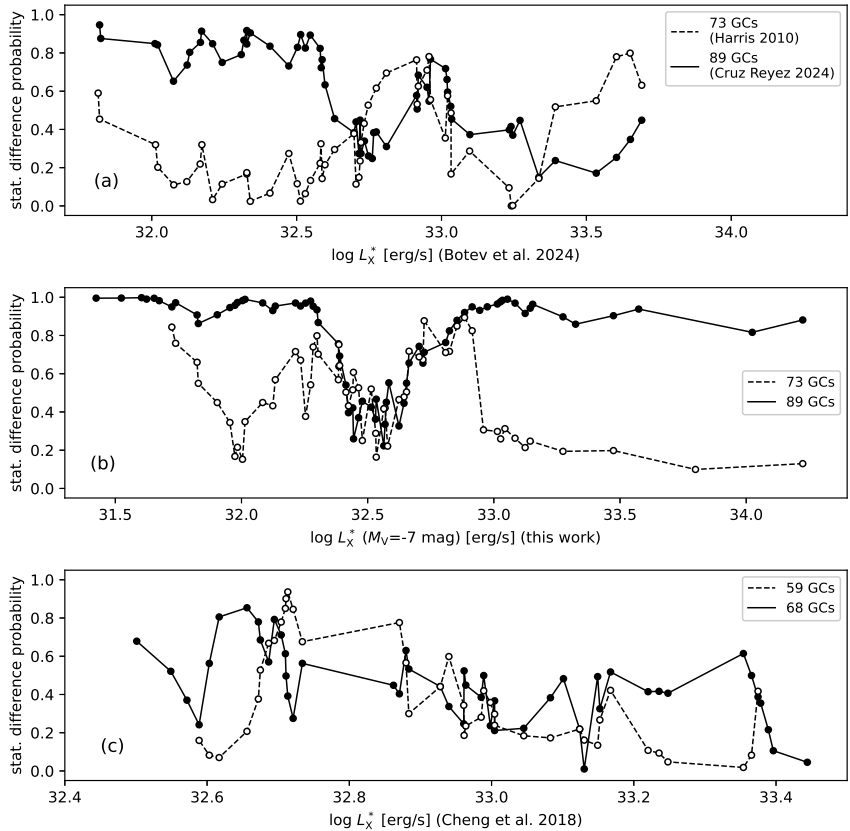

**Fig. 5.** K–S test probability for statistically different distributions of ellipticity as a function of the dividing X-ray luminosity threshold $L_X^*$. In the upper panel (a) the threshold $L_X^*$ follows the X-ray GCs luminosities $L_X([0.5–7]$ keV) within two core radii adopted by B24, in the middle panel (b) – the values, corresponding to the $L_X(M_V = -7$ mag) dividing lines in Fig. 3, and in the bottom panel (c) – the values of $L_X([0.5–8]$ keV) within half light radius adopted by C18. Everywhere, each K–S tests results are marked with open circles connected with a dashed line, when H10 ellipticities are used, or filled circles linked with solid line, when ellipticities CR24 are utilized. Total number of GCs with available both X-ray and ellipticity data are given in the embedded rectangles in each panel.

In Fig. 5, the upper (a), middle (b) and bottom (c) panels correspond to K–S results based on different X-ray luminosities: $L_X([0.5–7]$ keV) within two core radii adopted by B24, $L_X(M_V = -7$ mag) dividing lines in Fig. 3 (this work) and $L_X([0.5–8]$ keV) within half light radius adopted by C18, respectively. In general, the results are quite different, which can be explained both by the weak correlation between the two systems of ellipticities used and by the addition of new data on the ellipticities of 16 clusters in CR24 not previously measured in H10. Analogous to optical luminosity, it may be suggested that a threshold for the X-ray luminosity of globular clusters might separate them into samples with different distributions of their ellipticities. Since the brightest luminosities are rarer, this is expected to occur first at the X-ray luminosities higher side. Indeed, B24 postulated a threshold of $L_{\rm X}^* = 10^{33.65}$ erg s$^{-1}$ to distinguish between 11 clusters with high and 61 clusters with low X-ray luminosity, but the statistical difference probability found was as low as 80%. That result is reproduced by a dashed line in Fig. 5a, but not confirmed when the ellipticities of CR24 are used (see the solid line in the same panel). There, the probability fluctuates around the same confidence levels but for $L_{\rm X}^* \lesssim 10^{32.55}$ erg s$^{-1}$. As seen in Fig. 5a and Fig. 5c, a varying X-ray luminosity threshold $L_{\rm X}^*$ cannot provide significant difference probability at levels greater than 95%, either within the limits of $L_{\rm X}([0.5–7]$ keV) or within the limits of $L_{\rm X}([0.5–8]$ keV).

However, this becomes possible (see Fig. 5b) when the varying threshold $L_{\rm X}^*$ is defined on the log $L_{\rm X}$–$M_{\rm V}$ as the value of $L_{\rm X}$ at $M_{\rm V} = -7$ mag for a series of parallel lines leaving the different number of GCs above and below them. Thus, the threshold $L_{\rm X}^*$ accounts for both X-ray and optical luminosity, and the results are more or less consistent for the two sets of ellipticities. Of course, due to their incompleteness, the H10 data should be taken only for comparison.

The main discrepancies occur for $L_{\rm X}^* \gtrsim 10^{33}$ erg s$^{-1}$. B24 found that 11 clusters with X-ray luminosity higher than $L_{\rm X}^* \gtrsim 10^{33.65}$ erg s$^{-1}$ have an average ellipticity of 0.054, which is 0.030 less than the ellipticity of 61 clusters with X-ray luminosity lower than the limit. They also mentioned that eight of the 10 brightest X-ray Milky way GCs ($L_{\rm X} > 10^{34}$ erg s$^{-1}$) were lacking measured ellipticity and that gap was filled by CR24 which changes the picture completely.

Now, 20 clusters with X-ray luminosity higher than $L_{\rm X}^* \gtrsim 10^{33.05}$ erg s$^{-1}$ have an average ellipticity of 0.106, which is 0.049 greater than the ellipticity of 69 clusters with X-ray luminosity lower than that threshold. As the ellipticity is intricately related to rotational velocity (Bianchini et al., 2013; Fabricius et al., 2014), we suggest a possible link between the cluster angular momentum and the rate or X-ray binary formation. The physical processes leading to such a relation remain a matter of discussion. In this framework, clusters with significant internal rotation may display higher ellipticity and velocity dispersion, which together facilitate the dynamical formation of close binaries and, consequently, enhance the abundance of X-ray sources.

The threshold must be as low as $L_{\rm X}^* \gtrsim 10^{32.01}$ erg s$^{-1}$ in order for 64 clusters with higher X-ray luminosity to have an average ellipticity of 0.065, lower than the average ellipticity of 0.077 of 25 clusters below that threshold. Let us note that, due to the improved statistics, K–S test results testify for stable and significant difference probabilities for the two samples of ellipticity distributions within wide ranges of $L_{\rm X}^*$ rather than fluctuating probabilities even when the change in the number of accounted ellipticities is ±1. Between the threshold interval $L_{\rm X}^* = 10^{32.4} – 10^{32.6}$ erg s$^{-1}$, the statistical difference probability drops to ∼ 40%. We summarize the extreme cases of K–S probability $P_{\rm KS}$ for statistically different ellipticity $\epsilon$ distributions for sample pairs divided by a fixed threshold $L_{\rm X}^*$ in Table 3, together with sample statistics, average ellipticity values $\langle \epsilon \rangle$ and their standard deviations $\sigma$. We also add there the probabilities $P_{\rm ave}$ and $P_{\rm med}$ for a significantly different average and significantly different median of the same pairs of samples divided by the same threshold $L_X^*$.

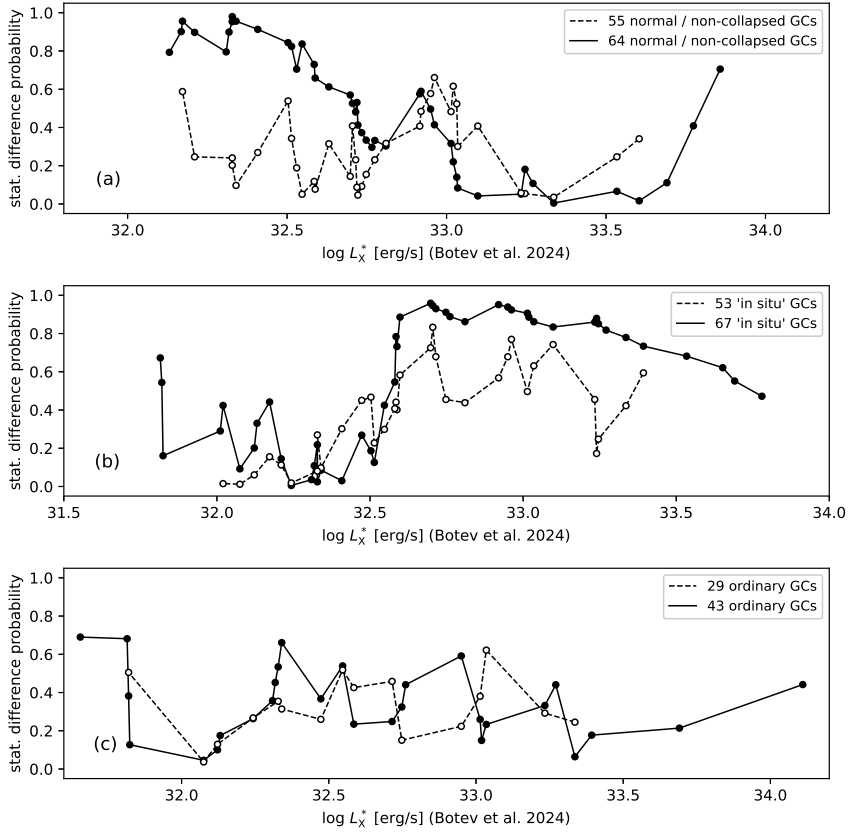

**Fig. 6.** K–S test probability for statistically different distributions of ellipticity as a function of the dividing X-ray luminosity threshold $L_X^*$. On all panels the threshold $L_X^*$ follows the X-ray GCs luminosities $L_X([0.5–7]\,\mathrm{keV})$ within two core radii adopted by B24. Three additional selection criteria were applied separately to data, used in samples 1 and sample 2 (see Table 1 and Table 2), namely 'only non-collapsed GCs' in the upper panel, 'only 'in situ' GCs' in the middle panel, and 'only ordinary GCs'. The designations of the open and filled circles, as well as the dashed and solid lines, are the same as those used in Fig. 5. Total number of GCs with available both X-ray and ellipticity data are given in the embedded rectangles in each panel.

In Fig. 6, the upper (a), middle (b) and bottom (c) panels correspond to K–S results based on one and the same X-ray luminosity, $L_X([0.5–7]\,\mathrm{keV})$ within two core radii adopted by B24 but for three different GCs types, respectively, 'non-collapsed' as classified in H10, 'in-situ' according to Belokurov & Kravtsov (2024) and 'ordinary', which do not posses any peculiarity like tails, tidal distortion etc., reported either in Grillmair et al. (1995) or Lehmann & Scholz (1997), or Leon et al. (2000), or Chen & Chen (2010). The implementation of such selection criteria leads to less numerous by $\sim 30\%$ samples 7–10 and even by $\sim 60\%$ samples 11–12, when only 'ordinary' GCs are considered (see Table 2).

As seen in Fig. 6c, for the latter type of GCs no certain conclusion can be drawn about any significant statistical difference or identity within the whole range of the dividing threshold $L_X^*$. Howewer, significant statistical differences occur (see Figs. 6a and 6b) for the 'non-collapsed' GCs when $L_X^* \lesssim 10^{32.4}$ erg s$^{-1}$ and contrary for the 'in situ' GCs when $L_X^* \gtrsim 10^{32.6}$ erg s$^{-1}$. Now, readdressing Fig. 5b, one may speculate that the left wing of significant probability difference is caused by predominance of 'non-collapsed' GC and the right wing – by the predominance of 'in situ' GC. Indeed, this hypothesis is supported not only by the high K–S corresponding probabilities 98.0% and 95.1%, reported in Table 3, but also by the fact that the average ellipticity for 'non-collapsed' GCs is 0.085 for $L_X^* < 10^{32.33}$ erg s$^{-1}$ against 0.064 for $L_X^* \geq 10^{32.33}$ erg s$^{-1}$ while for the 'in situ' GCs the average ellipticity is 0.067 for $L_X^* < 10^{32.90}$ erg s$^{-1}$ against 0.082 for $L_X^* \geq 10^{32.90}$ erg s$^{-1}$.

**Table 3.** The extreme cases of probability (significance level) in % for statistically different ellipticity $\epsilon$ distributions of two samples from CR24 and divided by a fixed threshold $L_X^*$. Columns are as follows: (1) log $L_X^*$ logarithm of the fixed X-ray luminosity threshold, (2) $n_1$ the number of GCs with lower $L_X$ than the threshold, (3) $\langle \epsilon \rangle \pm \sigma_\epsilon$ mean ellipticity for $n_1$ with its standard deviation, (4) $n_2$ the number of GCs with higher $L_X$ than the threshold, (5) $\langle \epsilon \rangle \pm \sigma_\epsilon$ mean ellipticity for $n_2$ with its standard deviation, (6) $P_{ave}$ probability for a significantly different average for the sample pair, (7) $P_{med}$ probability for a significantly different median for the sample pair, (8) $P_{KS}$ level of significance for statistically different ellipticity $\epsilon$ distributions according to performed K–S tests, (9) Shown in figure No.

| log $L_X^*$ erg s$^{-1}$ | $n_1(L_X < L_X^*)$ | $\langle \epsilon \rangle \pm \sigma_\epsilon$ | $n_2(L_X \geq L_X^*)$ | $\langle \epsilon \rangle \pm \sigma_\epsilon$ | $P_{ave}$ % | $P_{med}$ % | $P_{KS}$ % | Fig. No. |
|---|---|---|---|---|---|---|---|---|
| (1) | (2) | (3) | (4) | (5) | (6) | (7) | (8) | (9) |
| 33.05 | 69 | 0.057±0.041 | 20 | 0.106±0.077 | 100.0 | 98.9 | 99.0 | 5b |
| 32.56 | 48 | 0.062±0.046 | 41 | 0.075±0.063 | 75.1 | 33.7 | 22.3 | 5b |
| 32.01 | 25 | 0.077±0.054 | 64 | 0.065±0.055 | 65.0 | 95.5 | 98.9 | 5b |
| 33.32 | 49 | 0.061±0.048 | 14 | 0.068±0.060 | 34.3 | 17.0 | 0.5 | 6a |
| 32.33 | 17 | 0.085±0.063 | 47 | 0.054±0.043 | 97.2 | 98.2 | 98.0 | 6a |
| 32.90 | 41 | 0.067±0.056 | 26 | 0.082±0.065 | 67.4 | 73.7 | 95.1 | 6b |
| 32.24 | 20 | 0.073±0.062 | 47 | 0.073±0.059 | 3.9 | 10.4 | 0.5 | 6b |

## Conclusion

Based on the results presented in this paper the following conclusions can be drawn:

– No statistically significant correlation was identified between the ellipticity $\epsilon$ and the X-ray luminosity $L_X$, the absolute magnitude $M_V$, or the GC mass-to-$L_X$ ratio.

- Based on 70 K–S test results, statistically significant differences in the ellipticity distributions of Milky Way GCs were identified as a function of their X-ray luminosity. This was achieved through the use of homogeneous ellipticity data from CR24 and X-ray luminosities from B24, combined with a varying threshold $L_X^*$ that incorporates both X-ray and optical luminosity.
- Globular clusters with X-ray luminosity above the threshold $L_X^* = 10^{33.05}$ erg s$^{-1}$ exhibit a mean ellipticity of 0.106, which is significantly higher than the mean ellipticity of 0.057 for clusters below this threshold. This result contrasts with earlier conclusions based solely on optical data, where more luminous clusters were found to be rounder.
- At a lower threshold of $L_X^* = 10^{32.01}$ erg s$^{-1}$, the trend reverses: clusters above the threshold have a lower mean ellipticity (0.065) compared to those below it (0.077). This indicates a non-monotonic dependence of ellipticity on X-ray luminosity, sensitive to the choice of threshold.
- The most robust and repeatable K–S test results were obtained when the threshold $L_X^*$ was defined as the X-ray luminosity at $M_V = -7$ mag along a family of dividing lines in the log $L_X$ vs. $M_V$ plane. This approach yielded consistent results across different ellipticity datasets and sample selections.
- Globular clusters characterized by substantial internal rotation are expected to exhibit increased ellipticity and enhanced velocity dispersion, conditions that are favorable to the dynamical formation of close binary systems. These binaries, in turn, are efficient progenitors of X-ray sources, thereby potentially elevating the cluster's overall X-ray luminosity.

## Acknowledgements

This study is financed by Scientific research fund of Sofia University "St. Kliment Ohridski", contract 80-10-34/23.05.2025.

Petko Nedialkov also acknowledges a partial support by the European Union – NextGenerationEU, through the National Recovery and Resilience Plan of the Republic of Bulgaria, project SUMMIT BG-RRP-2.004-0008-C01.

We are grateful to the anonymous and well-intentioned referee for the constructive comments, which helped to improve the clarity and overall quality of this paper.